\documentclass[prd,twocolumn,showpacs,preprintnumbers,superscriptaddress,floatfix,nofootinbib]{revtex4-1}
\usepackage{amsfonts}
\usepackage{graphicx,,booktabs}
\usepackage{amssymb}
\usepackage{amsmath,txfonts}
\usepackage{graphicx}
\usepackage{dcolumn}
\usepackage{color}
\usepackage{bm}
\usepackage[colorlinks,
            citecolor=blue,
            anchorcolor=green,
            menucolor=orange,
            linkcolor=red,
            filecolor=red,
            runcolor=pink,
            urlcolor=blue,
            frenchlinks=red]{hyperref}

              \allowdisplaybreaks[4]

\begin{document}

\title{\boldmath Possibility to study pentaquark states $%
P_{c}(4312)$, $P_{c}(4440)$, and $P_{c}(4457)$ in the reaction $\gamma p\rightarrow
J/\psi p$}
\author{Xiao-Yun Wang}
\thanks{xywang01@outlook.com}
\affiliation{Department of physics, Lanzhou University of Technology,
Lanzhou 730050, China}
\author{Xu-Rong Chen}
\thanks{xchen@impcas.ac.cn}
\affiliation{Institute of Modern Physics, Chinese Academy of Sciences, Lanzhou 730000,
China}
\author{Jun He}
\thanks{Corresponding author : junhe@njnu.edu.cn}
\affiliation{Department of  Physics and Institute of Theoretical Physics, Nanjing Normal University,
Nanjing, Jiangsu 210097, China}
\begin{abstract}
Inspired by the observation of the pentaquark states $P_{c}(4312),P_{c}(4440)$
and $P_{c}(4457)$ at LHCb, photoproduction of these three $P_{c}$ states via
the interaction $\gamma p\rightarrow J/\psi p$ is investigated in an effective
Lagrangian approach. The $t$-channel Pomeron exchange diffractive process is
considered as the main background for the $J/\psi $ photoproduction. The
numerical results show that the theoretical cross section, which is
calculated by assuming a branching ratio $Br[P_{c}\rightarrow J/\psi
p]\simeq 3\%$, is consistent with the existing experimental data of the $%
\gamma p\rightarrow J/\psi p$ process. With such a branching ratio, if
experimental precision reaches 0.1 nb within a bin of 100 MeV for photon
energy, two peaks are expected to be obviously observed in the $J/\psi $
photoproduction. To observe the two-peak structure from $P_{c}(4440)$ and $%
P_{c}(4457)$, higher precision, about 0.1nb/10 MeV, is required to
distinguish two close pentaquarks. If the physical branching ratio is
larger, the requirement of experimental precision will be reduced. The
differential cross sections for reaction $\gamma p\rightarrow J/\psi p$ are
also present. It is found that the $t$-channel Pomeron exchange provides a
sharp increase at extreme forward angles and gives a sizable contribution at
most energy points, while the contributions from the $s$-channel $P_{c}$
exchanges play important roles at threshold energies. The experimental
measurement of the $\gamma p\rightarrow J/\psi p$ process in the
near-threshold energy region around $E_{\gamma }\simeq 9.4-10.5$ GeV is
suggested, and is accessible at CEBAF@JLab and COMPASS.
\end{abstract}

\maketitle

\section{Introduction}

As of now, many exotic hadrons have been observed and listed in the Review
of Particle Physics (PDG)~\cite{Tanabashi:2018oca}. However, the internal
structure of these exotic hadrons is still a confusing problem. The
pentaquark is a type of important exotic hadron. In 2015, LHCb reported
their observation of two pentaquark candidates, $P_{c}(4450)$ and $%
P_{c}(4380)$~\cite{Aaij:2015tga}. Very recently, an updated result was
reported: three narrow pentaquark states, $P_{c}(4312),P_{c}(4440)$, and $%
P_{c}(4457)$, were observed in the $J/\psi p$ invariant mass spectrum of the
$\Lambda _{b}\rightarrow J/\psi pK$ decays~\cite{Aaij:2019vzc}. It is
interesting to see that the $P_{c}(4450)$ formerly reported by the LHCb
Collaboration \cite{Aaij:2015tga} splits into two peaks, $P_{c}(4440)$ and $%
P_{c}(4457)$, based on more accumulated data. The masses and widths of these
three $P_{c}$ states were measured~\cite{Aaij:2019vzc}:%
\begin{eqnarray}
P_{c}^{+}(4312):M &=&4311.9\pm 0.7_{-0.6}^{+6.8}\mbox{ MeV}\,,  \notag \\
\Gamma  &=&9.8\pm 2.7_{-4.5}^{+3.7}\mbox{ MeV}\,,  \notag \\
P_{c}^{+}(4440):M &=&4440.3\pm 1.3_{-4.7}^{+4.1}\mbox{ MeV}\,,  \notag \\
\Gamma  &=&20.6\pm 4.9_{-10.1}^{+8.7}\mbox{ MeV}\,,  \notag \\
P_{c}^{+}(4457):M &=&4457.3\pm 0.6_{-1.7}^{+4.1}\mbox{ MeV}\,,  \notag \\
\Gamma  &=&6.4\pm 2.0_{-1.9}^{+5.7}\mbox{ MeV}\,.  \label{experiment}
\end{eqnarray}

The nature of these three $P_{c}$ states attracts much attention and has
been studied by many theoretical models~\cite%
{Chen:2019asm,Chen:2019bip,Liu:2019tjn,He:2019ify,Huang:2019jlf,Ali:2019npk,Xiao:2019mvs,Shimizu:2019ptd,Guo:2019kdc,Xiao:2019aya,Guo:2019fdo,Cao:2019kst}%
. Among these theoretical investigations, the spin parities of these $P_{c}$
states were also predicted. It can be seen that these three $P_{c}$ states
are quite narrow and can be clearly seen in the $J/\psi p$ invariant mass
spectrum. Moreover, closeness of the $\Sigma _{c}\bar{D}$ or $\Sigma _{c}%
\bar{D}^{\ast }$ thresholds to these three structures suggests that the $%
\Sigma _{c}\bar{D}$ and $\Sigma _{c}\bar{D}^{\ast }$ interactions play
important roles in the dynamics of these $P_{c}$ states. Accordingly, one
notices that many of theoretical studies suggested that the $P_{c}(4312)$
can be assigned as an S-wave $\Sigma _{c}\bar{D}$ bound state with spin parity 
$1/2^{-}$ and the $P_{c}(4440)$ and $P_{c}(4457)$ as S-wave $\Sigma
_{c}\bar{D}^{\ast }$ bound states with spin parities $1/2^{-}$ and $3/2^{-}$%
, respectively~\cite%
{Chen:2019asm,Liu:2019tjn,He:2019ify,Huang:2019jlf,Xiao:2019mvs,Xiao:2019aya}.  At the hadronic level, the $P_c$ states as molecular states are  generated from the $\Sigma^{(*)} \bar{D}^{(*)}$  interaction  in the one-boson-exchange model in which pseudoscalar ($\pi$, $\eta$), vector ($\rho$, $\omega$)  and scalar $\sigma$ exchanges are considered~\cite{Chen:2019asm,He:2019ify,He:2015cea,Yang:2011wz}.  The existence of such molecular states is also confirmed by the calculation in the constituent quark model with the meson-exchange mechanism at the quark level~\cite{Liu:2019tjn,Huang:2019jlf}.

Up to now, these hidden-charm pentaquarks were only observed in the $\Lambda
_{b}$ decay at LHCb. Production of the pentaquarks in other ways is very
helpful to obtain the definite evidence for their nature as genuine states.
Before  LHCb's observation of $P_{c}(4450)$ and $%
P_{c}(4380)$ , the production of hidden-charm pentaquark in the $p\bar{p}%
\rightarrow p\bar{p}J/\psi $ process was proposed in the article in which the
hidden-charm pentaquark was predicted~\cite{Wu:2010jy}. The photoproduction
of the hidden-charm pentaquark was first suggested to be applied at
Jefferson Laboratary in Ref.~\cite{Huang:2013mua}. After the LHCb experiment, more
works about the production of the $P_{c}(4450)$ and $P_{c}(4380)$ appeared~%
\cite{Wang:2015jsa,Lu:2015fva,Garzon:2015zva,Karliner:2015voa}. In
particular an experimental proposal to search for the $P_{c}(4450)$ in  $%
J/\psi $ photoproduction was put forward to be performed in Hall C at
Jefferson Lab~\cite{Meziani:2016lhg}. At present, there are already some
experimental data~\cite{cornell75,SLAC75,SLAC76} for the  reaction $\gamma
p\rightarrow J/\psi p$. One notes that there exist large
uncertainties in existing experimental data~\cite{cornell75,SLAC75,SLAC76}
in the near-threshold energy region where the $P_{c}$ states live. The
high-luminosity detectors at Jefferson Laboratary will produce high-precision data
in future experiments. One can expect these experiments to provide an
opportunity to study the hidden-charm pentaquarks in photoproduction.

In the new LHCb results, the $P_{c}(4450)$ splits into two states, $%
P_{c}(4440)$ and $P_{c}(4457)$. These two states are quite narrower than the $%
P_{c}(4450)$, which will affect previous predictions about the
photoproduction of the $P_{c}$ states. Moreover, a new pentaquark $%
P_{c}(4312)$ was observed. Hence, it is interesting to restudy the
photoproduction of the pentaquarks based on the new LHCb results~\cite%
{Aaij:2019vzc}. In this paper, within the framework of an effective
Lagrangian approach, the photoproduction of three $P_{c}$ states via  reaction $\gamma
p\rightarrow J/\psi p$ is investigated. The $t$-channel Pomeron
exchange diffractive process is considered as the main background.

This paper is organized as follows. After the Introduction, we present the
formalism including Lagrangians and amplitudes for the $\gamma p\rightarrow
J/\psi p$ process in Sec.~II. The numerical results of the cross section
follow in Sec.~III. Finally, the paper{\ ends} with a brief summary.

\section{Formalism}

\begin{figure}[tbph]
\begin{center}
\includegraphics[scale=0.52]{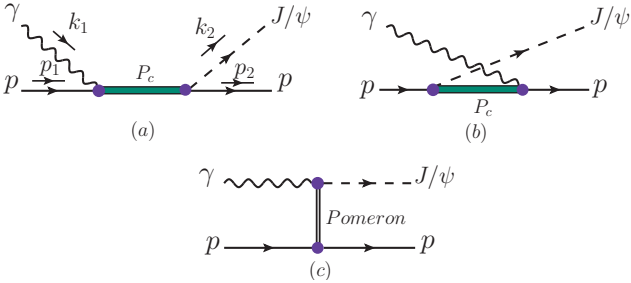}
\end{center}
\caption{Feynman diagrams for the reaction $\protect\gamma %
p\rightarrow J/\protect\psi p$ .}
\label{Fig: Feynman}
\end{figure}

The basic tree-level Feynman diagrams of the 
reactions $\gamma p\rightarrow J/\psi p$ are illustrated in Fig.~\ref{Fig: Feynman} in which the pentaquark $%
P_{c}$ candidates are produced through $s$ and $u$ channel. The background
contribution is mainly from the $t$-channel Pomeron exchange, as depicted in
Fig.~\ref{Fig: Feynman}(c). Considering the off-shell effect of the
intermediate $P_{c}$ states, the $u$-channel contribution will be neglected
in our calculation.

\subsection{Lagrangians for $P_{c}$ photoproduction}

At present, the spin-parity quantum numbers of these $P_{c}$ states were not
determined experimentally. In this work, the theoretical prediction of $%
P_{c}(4312)$ with $J^{P}=1/2^{-}$, $P_{c}(4440)$ with $J^{P}=1/2^{-}$ and $%
P_{c}(4457)$ with $J^{P}=3/2^{-}$ are taken in our calculation as suggested
in Refs.~\cite%
{Chen:2019asm,Liu:2019tjn,He:2019ify,Huang:2019jlf,Xiao:2019mvs,Xiao:2019aya}
. For describing the $P_{c}$ photoproduction process, 
Lagrangians are needed \cite%
{Oh:2007jd,Wan:2015gsl,Wang:2015hfm,Kim:2011rm,Wang:2015jsa},
\begin{eqnarray}
\mathcal{L}_{\gamma NP_{c}}^{1/2^{-}} &=&\frac{eh}{2m_{N}}\bar{N}\sigma
_{\mu \nu }\partial ^{\nu }A^{\mu }P_{c}+{\rm H.c.}, \\
\mathcal{L}_{P_{c}\psi N}^{1/2^{-}} &=&g_{P_{c}\psi N}^{1/2^{-}}\bar{N}%
\gamma _{5}\gamma _{\mu }P_{c}\psi ^{\mu }+{\rm H.c.}, \\
\mathcal{L}_{\gamma NP_{c}}^{3/2^{-}} &=&e(\frac{ih_{1}}{2m_{N}}\bar{N}%
\gamma ^{\nu }-\frac{h_{2}}{(2m_{N})^{2}}\partial ^{\nu }\bar{N})  \notag \\
&&\times F_{\mu \nu }P_{c}^{\mu }+{\rm H.c.},  \label{L1} \\
\mathcal{L}_{P_{c}\psi N}^{3/2^{-}} &=&\frac{-ig_{P_{c}\psi N}^{3/2^{-}}}{%
2m_{N}}\bar{N}\gamma _{\nu }\psi ^{\mu \nu }P_{c\mu }-\frac{g_{2}}{%
(2m_{N})^{2}}\partial _{\nu }\bar{N}\psi ^{\mu \nu }P_{c\mu }  \notag \\
&&+\frac{g_{2}}{(2m_{N})^{2}}\bar{N}\partial _{\nu }\psi ^{\mu \nu }P_{c\mu
}+{\rm H.c.},  \label{L2}
\end{eqnarray}%
where $F_{\mu \nu }=\partial _{\mu }A_{\nu }-\partial _{\nu }A_{\mu }$ and $%
{N}$, $A$, $P_{c}$, and $\psi $ are the nucleon, photon, $P_{c}$ state, and $%
J/\psi $ meson fields, respectively.

Since the momenta of the final states $J/\psi N$ are fairly small compared
with the nucleon mass, the higher partial wave terms in Eqs.~(\ref{L1}) and (%
\ref{L2}) will be neglected in the following, calculated as done in Ref.~\cite%
{Wang:2015jsa}.

The value of $g_{P_{c}\psi N}^{1/2^{-}}$ and $g_{P_{c}\psi N}^{3/2^{-}}$ can
be determined from the decay width%
\begin{eqnarray}
\Gamma _{P_{c}\rightarrow \psi N}^{1/2^-}&=\frac{|\vec{p}_{\psi }^{~\mathrm{%
c.m.}}|}{16\pi m_{P_{c}}^{2}}\left\vert \mathcal{M}_{P_{c}\rightarrow \psi
N}^{1/2^-}\right\vert ^{2}, \\
\Gamma _{P_{c}\rightarrow \psi N}^{3/2^{-}}&=\frac{|\vec{p}_{\psi }^{~%
\mathrm{c.m.}}|}{32\pi m_{P_{c}}^{2}}\left\vert \mathcal{M}%
_{P_{c}\rightarrow \psi N}^{3/2^{-}}\right\vert ^{2},
\end{eqnarray}

with

\begin{equation}
|\vec{p}_{\psi }^{~\mathrm{c.m.}}|=\frac{\lambda (m_{P_{c}}^{2},m_{\psi
}^{2},m_{N}^{2})}{2m_{P_{c}}},
\end{equation}%
where $\lambda $ is the K\"{a}llen function with $\lambda (x,y,z)\equiv
\sqrt{(x-y-z)^{2}-4yz}$. The $m_{P_{c}}$, $m_{\psi }$, and $m_{N}$ are the
masses of $P_{c}$, $J/\psi $, and the nucleon. $\mathcal{M}_{P_{c}\rightarrow
\psi N}^{1/2^{-}}$ and $\mathcal{M}_{P_{c}\rightarrow \psi N}^{3/2^{-}}$ are
the decay amplitudes.

For the electromagnetic (EM) coupling $eh$ related to the $\gamma NP_{c}$
vertex, its value can be obtained from the strong coupling constant $%
g_{P_{c}\psi N}$ within the vector meson dominance (VMD) mechanism \cite%
{tb65,tb69,tb79}. In the frame of the VMD mechanism, the EM coupling
constants $eh$ and $eh_{1}$ are related to the coupling constants $%
g_{P_{c}\psi N}^{1/2^{-}}$ and $g_{P_{c}\psi N}^{3/2^{-}}$ as%
\begin{eqnarray}
eh &=&g_{P_{c}\psi N}^{1/2^{-}}\frac{e}{f_{\psi }}\frac{2m_{N}}{%
(m_{P_{c}}^{2}-m_{N}^{2})m_{\psi }}  \notag \\
&&\times \sqrt{m_{\psi
}^{2}(m_{N}^{2}+4m_{N}m_{P_{c}}+m_{P_{c}}^{2})+(m_{P_{c}}^{2}-m_{N}^{2})^{2}}%
, \\
eh_{1} &=&g_{P_{c}\psi N}^{3/2^{-}}\frac{e}{f_{\psi }}\frac{%
2m_{N}(m_{N}+m_{P_{c}})}{(m_{P_{c}}^{2}-m_{N}^{2})m_{\psi }}  \notag \\
&&\times \sqrt{\frac{6m_{\psi
}^{2}m_{P_{c}}^{2}+m_{N}^{4}-2m_{N}^{2}m_{P_{c}}^{2}+m_{P_{c}}^{4}}{%
3m_{P_{c}}^{2}+m_{N}^{2}}}.
\end{eqnarray}%
\ \ \

The Lagrangian depicting the coupling of the meson $J/\psi $ with a photon
reads as%
\begin{equation}
\mathcal{L}_{J/\psi \gamma }=-\frac{em_{\psi }^{2}}{f_{\psi }}V_{\mu }A^{\mu
},
\end{equation}%
where $f_{\psi }$ is the $J/\psi $ decay constant. Thus one
gets the expression for the $J/\psi \rightarrow e^{+}e^{-}$ decay,%
\begin{equation}
\Gamma _{J/\psi \rightarrow e^{+}e^{-}}=\left( \frac{e}{f_{\psi }}\right)
^{2}\frac{8\alpha \left\vert \vec{p}_{e}^{~\mathrm{c.m.}}\right\vert ^{3}}{%
3m_{\psi }^{2}},
\end{equation}%
where $\vec{p}_{e}^{~\mathrm{c.m.}}$ denotes the 3-momentum of an
electron in the rest frame of the $J/\psi $ meson. The $\alpha =e^{2}/4\hbar
c=1/137$ is the electromagnetic fine structure constant. With the partial
decay width of $J/\psi \rightarrow e^{+}e^{-}$ \cite{Tanabashi:2018oca},%
\begin{equation}
\Gamma _{J/\psi \rightarrow e^{+}e^{-}}\simeq 5.547\text{ keV,}
\end{equation}%
one gets $e/f_{\psi }\simeq 0.027$.

Finally, one obtains the EM couplings related to the $\gamma NP_{c}$ vertices
and coupling constants $g_{P_{c}\psi N}$ from partial decay widths $\Gamma
_{P_{c}\rightarrow \psi N}$ with different $J^{P}$ assignments of the $P_{c}$
states. The obtained coupling constants are listed in Tables~\ref{tab1} and \ref%
{tab2} by assuming that the $J/\psi p$ channel accounts for $3\%$ and $10\%$
of total widths of the $P_{c}$ states, respectively. \renewcommand%
\tabcolsep{0.7cm} \renewcommand{\arraystretch}{2}
\begin{table}[tbph]
\caption{The values of coupling constants by assuming the $J/\protect\psi p$
channel accounts for $3\%$ of total widths of the $P_{c}$ states. }%
\begin{tabular}{ccc}
\hline\hline
States & $g_{P_{c}\psi N}$ & $eh$ or $eh_{1}$ \\ \hline
$P_{c}(4312)$ $(J^{P}=\frac{1}{2}^{-})$ & 0.06 & 0.0014 \\
$P_{c}(4440)$ $(J^{P}=\frac{1}{2}^{-})$ & 0.08 & 0.0018 \\
$P_{c}(4457)$ $(J^{P}=\frac{3}{2}^{-})$ & 0.036 & 0.0008 \\ \hline\hline
\end{tabular}%
\label{tab1}
\end{table}

\begin{table}[tbph]
\caption{The values of coupling constants by assuming the $J/\protect\psi p$
channel accounts for $10\%$ of total widths of the $P_{c}$ states. }%
\begin{tabular}{ccc}
\hline\hline
States & $g_{P_{c}\psi N}$ & $eh$ or $eh_{1}$ \\ \hline
$P_{c}(4312)$ $(J^{P}=\frac{1}{2}^{-})$ & 0.11 & 0.0026 \\
$P_{c}(4440)$ $(J^{P}=\frac{1}{2}^{-})$ & 0.14 & 0.0032 \\
$P_{c}(4457)$ $(J^{P}=\frac{3}{2}^{-})$ & 0.07 & 0.0016 \\ \hline\hline
\end{tabular}%
\label{tab2}
\end{table}

\subsection{Pomeron exchange}

Since the Pomeron can mediate the long-range interaction between a confined
quark and a nucleon \cite%
{lxh08,Wang:2015lwa,Donnachie:1987pu,Pichowsky:1996jx}, the $t$-channel
Pomeron exchange [as described in Fig. 1(c)] is considered  the main
background contribution to the $P_{c}(4312)$ photoproduction process. The
Pomeron exchange is expressed in terms of the quark loop coupling in the
vertices of $\mathbb{P}NN$ and $\gamma \mathbb{P}\psi $. The Pomeron-nucleon
coupling is written as \cite%
{lxh08,Wang:2015lwa,Donnachie:1987pu,Pichowsky:1996jx}%
\begin{equation}
\mathcal{F}_{\rho }(t)=\frac{3\beta _{0}(4m_{N}^{2}-2.8t)}{%
(4m_{N}^{2}-t)(1-t/0.7)^{2}}\gamma _{\rho }=F(t)\gamma _{\rho }\text{,}
\end{equation}%
where $t=q_{P}^{2}$ is the exchanged Pomeron momentum squared. $\beta
_{0}^{2}=4$ GeV$^{2}$ stands for the coupling constant between a single
Pomeron and a light constituent quark.

For the $\gamma \mathbb{P}\psi $ vertex, we have%
\begin{equation}
V_{\gamma \mathbb{P}\psi }=\frac{2\beta _{c}\times 4\mu _{0}^{2}}{(m_{\psi
}^{2}-t)(2\mu _{0}^{2}+m_{\psi }^{2}-t)}T_{\rho ,\mu \nu }\epsilon _{\psi
}^{\nu }\epsilon _{\gamma }^{\mu }\mathbb{P}^{\rho },
\end{equation}%
with%
\begin{equation*}
T^{\rho ,\mu \nu }=(k_{1}+k_{2})^{\rho }g^{\mu \nu }-2k_{1}^{\nu }g^{\rho
\mu }
\end{equation*}%
where $\beta _{c}^{2}=0.8$ GeV$^{2}$ is the effective coupling constant
between a Pomeron and a charm quark within the $J/\psi $ meson \cite%
{lxh08,Laget:1994ba}, while $\mu _{0}=1.2$ GeV$^{2}$ denotes a cutoff
parameter in the form factor of the Pomeron \cite%
{lxh08,Wang:2015lwa,Donnachie:1987pu,Pichowsky:1996jx}.

\subsection{Amplitudes}

According to the above Lagrangians, the scattering amplitude of the reaction $\gamma
p\rightarrow J/\psi p$ can be written as%
\begin{equation}
-i\mathcal{M}=\epsilon _{\psi }^{\nu }(k_{2})\bar{u}(p_{2})\mathcal{A}_{\mu
\nu }^{i}u(p_{1})\epsilon _{\gamma }^{\mu }(k_{1}),
\end{equation}%
where $u$ is the Dirac spinor of nucleon, and $\epsilon _{\psi }^{\nu }$ and
$\epsilon _{\gamma }^{\mu }$ are the polarization vector of $J/\psi $ meson
and photon, respectively.

The reduced amplitudes $\mathcal{A}_{\mu \nu }^{i}$ for the $s$ channel with
each $J^{P}$ assignment of $P_{c}$ state and the $t$ channel read
\begin{eqnarray}
\mathcal{A}_{\mu \nu }^{s(1/2^{-})} &=&\frac{eh}{2m_{N}}g_{P_{c}\psi N}%
\mathcal{F}(q^{2})\gamma _{5}\gamma _{\nu }\frac{(\rlap{$\slash$}q+m_{P_{c}})%
}{s-m_{P_{c}}^{2}+im_{P_{c}}\Gamma _{P_{c}}}  \notag \\
&&\times \gamma _{\mu }\rlap{$\slash$}k_{1},  \label{AmpT1} \\
\mathcal{A}_{\mu \nu }^{s(3/2^{-})} &=&\frac{eh_{1}}{2m_{N}}\frac{%
g_{P_{c}\psi N}}{2m_{N}}\mathcal{F}(q^{2})\gamma _{\sigma }(k_{2}^{\beta
}g^{\nu \sigma }-k_{2}^{\sigma }g^{\nu \beta })  \notag \\
&&\times \frac{(\rlap{$\slash$}q+m_{P_{c}})}{s-m_{P_{c}}^{2}+im_{P_{c}}%
\Gamma _{P_{c}}}\Delta _{\beta \alpha }\gamma _{\delta }(k_{1}^{\alpha
}g^{\mu \delta }-k_{1}^{\delta }g^{\alpha \mu }), \\
\mathcal{A}_{\mu \nu }^{t} &=&8\beta _{c}\mu _{0}^{2}\frac{\mathcal{F}_{\rho
}(t)G_{P}(s,t)}{(m_{\psi }^{2}-t)(2\mu _{0}^{2}+m_{\psi }^{2}-t)}T^{\rho
,\mu \nu },  \label{AmpT2}
\end{eqnarray}%
with
\begin{eqnarray}
\Delta _{\beta \alpha } &=&-g_{\beta \alpha }+\frac{1}{3}\gamma ^{\beta
}\gamma ^{\alpha }  \notag \\
&&+\frac{1}{3m_{P_{c}}}(\gamma ^{\beta }q^{\alpha }-\gamma ^{\alpha
}q^{\beta })+\frac{2}{3m_{P_{c}}^{2}}q^{\beta }q^{\alpha }, \\
G_{P}(s,t) &=&-i(\eta ^{\prime }s)^{\eta (t)-1}
\end{eqnarray}%
where $\eta (t)=1+\epsilon +\eta ^{\prime }t$ is the Pomeron Regge
trajectory, while the concrete values $\epsilon =0.08$ and $\eta ^{\prime
}=0.25$ GeV$^{-2}$ are adopted \cite%
{lxh08,Wang:2015lwa,Donnachie:1987pu,Pichowsky:1996jx}. Moreover, $%
s=(k_{1}+p_{1})^{2}$ and $t=(k_{1}-k_{2})^{2}$ are the Mandelstam variables.

For the $s$ channel with an intermediate $P_{c}$ state, one adopts a general
form factor to describe the size of hadrons as \cite%
{Wang:2015jsa,Wang:2017qcw,Wang:2015xwa}

\begin{equation}
\mathcal{F}(q^{2})=\frac{\Lambda ^{4}}{\Lambda ^{4}+(q^{2}-m_{P_{c}}^{2})^{2}%
}~,
\end{equation}%
where $q$ and $m_{P_{c}}$ are the 4-momentum and mass of the exchanged $%
P_{c}$ state, respectively. {For the heavier hadron production, the }typical
cutoff value $\Lambda =0.5$ GeV will be{\ taken }as used in Refs. \cite%
{Kim:2011rm,Wang:2015jsa}.

\section{Numerical results}

With the preparation in the previous section, the cross section of the  reaction $%
\gamma p\rightarrow J/\psi p$ can be calculated. The differential
cross section in the c.m. frame is written as
\begin{equation}
\frac{d\sigma }{d\cos \theta }=\frac{1}{32\pi s}\frac{\left\vert \vec{k}%
_{2}^{{~\mathrm{c.m.}}}\right\vert }{\left\vert \vec{k}_{1}^{{~\mathrm{c.m.}}%
}\right\vert }\left( \frac{1}{4}\sum\limits_{\lambda }\left\vert \mathcal{M}%
\right\vert ^{2}\right) ,
\end{equation}%
where $s=(k_{1}+p_{1})^{2}$ and $\theta $ denotes the angle of the outgoing
$J/\psi $ meson relative to photon beam direction in the c.m. frame. $\vec{k}%
_{1}^{{~\mathrm{c.m.}}}$ and $\vec{k}_{2}^{{~\mathrm{c.m.}}}$ are the
3-momenta of the initial photon beam and final $J/\psi $ meson,
respectively.

\begin{figure}[h!]
\begin{center}
\includegraphics[scale=0.41]{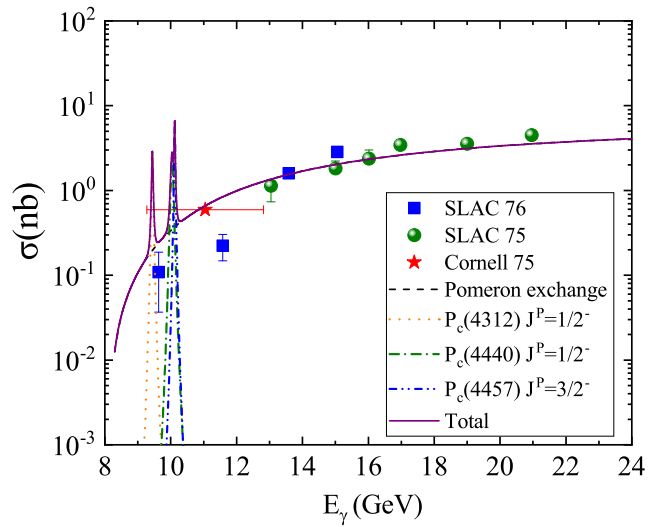}
\end{center}
\caption{Total cross section for the reaction $\protect\gamma %
p\rightarrow J/\protect\psi p$ by assuming branching ratio $%
Br[P_{c}\rightarrow J/\protect\psi p]\simeq 3\%$. The black dashed, orange
dotted, green dot-dashed, blue dash-double dotted, and violet solid lines
are for the Pomeron exchange, $P_{c}(4312),P_{c}(4440),P_{c}(4457)$, and
total contributions, respectively. The experimental data are from Refs.~%
\protect\cite{cornell75,SLAC75,SLAC76}.}
\label{total01}
\end{figure}

In Fig.~\ref{total01}, we present the total cross section for the  reaction $\gamma
p\rightarrow J/\psi p$ by assuming a branching ratio $%
Br[P_{c}\rightarrow J/\psi p]\simeq 3\%$ from threshold to 24 GeV of the
photon beam energy. It is found that the line shape of the total cross
section including both $P_{c}$ and Pomeron contributions goes up very
rapidly. The experimental data of the $\gamma p\rightarrow J/\psi p$ process
seem to be quite consistent with the total cross section. Moreover, the value
of the total cross section becomes larger and larger with the increase of the
beam energy up to 24 GeV. The monotonically increasing behavior should be
caused by the $t$-channel Pomeron exchange. Obviously, the contribution from
the Pomeron diffractive process is responsible for explaining the experimental
data points at high energies. The $P_{c}(4312)$ exhibits itself as an
independent peak, while the $P_{c}(4440)$ and $P_{c}(4457)$ are very close to
each other with the masses observed at LHCb.

\begin{figure}[h!]
\begin{center}
\includegraphics[scale=0.41]{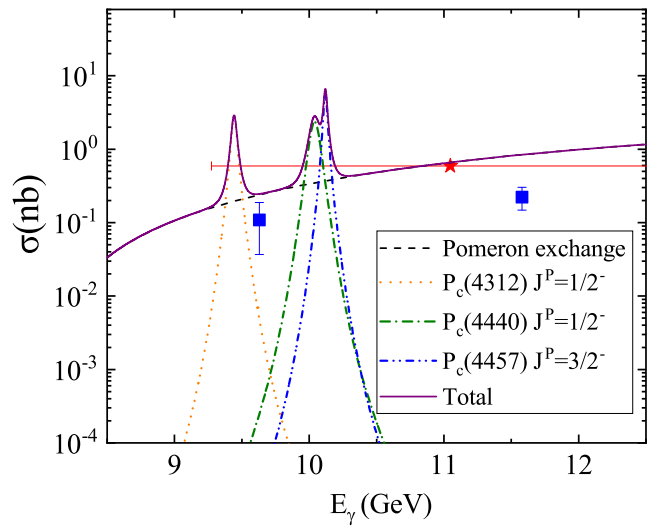}
\end{center}
\caption{Same as Fig. 2 except that the energy range is
reduced.}
\label{total02}
\end{figure}

To show the results for three pentaquarks more clearly, in Fig.~\ref{total02},
we give the same results as Fig.~\ref{total01} but with a reduced energy
range. The peak for the $P_{c}(4312)$ still stands independently. The $%
P_{c}(4450)$ and $P_{c}(4457)$ can be distinguished in the reduced energy
region. The results suggest that the $P_{c}(4312)$ can be observed within a
bin of 0.1 GeV. For  the two higher pentaquarks, $P_c(4440)$ and $P_c(4457)$, the  mass difference is only about 17 MeV, which  is comparable to the widths of these two pentaquarks. Hence, if we adopt a large bin, such as 0.1 GeV,  the $P_{c}(4440)$ and $P_{c}(4457)$ will exhibit  as
one resonance. The dip in the two-peak structure of these two states will disappear because the cross section should be averaged  in a bin.   To distinguish these two close states, we should choose several energy points between two peaks of $P_c(4440)$ and $P_c(4457)$,  which requires a
bin at least at an order of 10 MeV based on our calculation.  Our results suggest that the total
cross section is of order 1 nb. Hence, by assuming a branching ratio $%
Br[P_{c}\rightarrow J/\psi p]\simeq 3\%$, to observe the $P_{c}(4312)$, the
experimental precision should be 1nb/0.1 GeV. However, to observe the
two-peak structure from the $P_{c}(4440)$ and $P_{c}(4457)$, higher
precision, 0.1nb/10 MeV, is required based on our results.

\begin{figure}[h!]
\begin{center}
\includegraphics[scale=0.41]{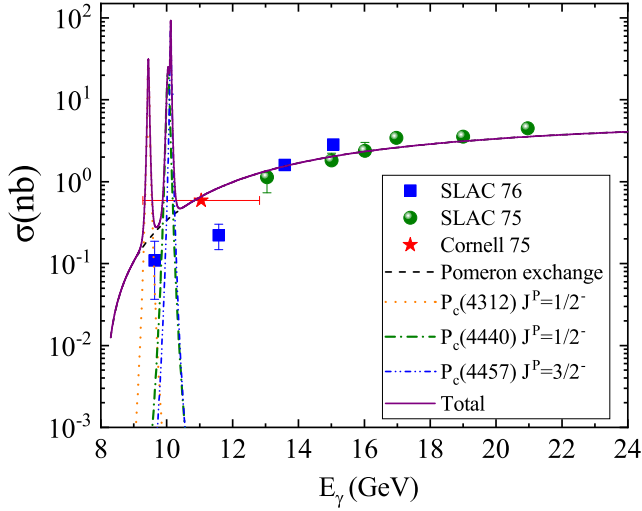}
\end{center}
\caption{Same as Fig. 2 except assuming a
branching ratio $Br[P_{c}\rightarrow J/\protect\psi p]\simeq 10\%.$}
\label{total03}
\end{figure}

In Fig.~\ref{total03}, we show the obtained total cross section of the $%
\gamma p\rightarrow J/\psi p$ process as a function of the photon beam
energy, where the result are calculated by assuming $Br[P_{c}\rightarrow
J/\psi p]\simeq 10\%$. It can be seen that the total cross section exhibits
three peaks near the threshold. In the near-threshold energy region, the
contributions from the $s$-channel $P_{c}$-state exchanges are at least an
order of magnitude higher than the cross section from the background from
the Pomeron exchange, which indicates that the signal can be clearly
distinguished from the background. Thus, the range $E_{\gamma }\simeq
9.4-10.5$ GeV should be the best energy window of searching for these $P_{c}$
states via the photoinduced process.

\begin{figure}[h!]
\centering
\includegraphics[scale=0.41]{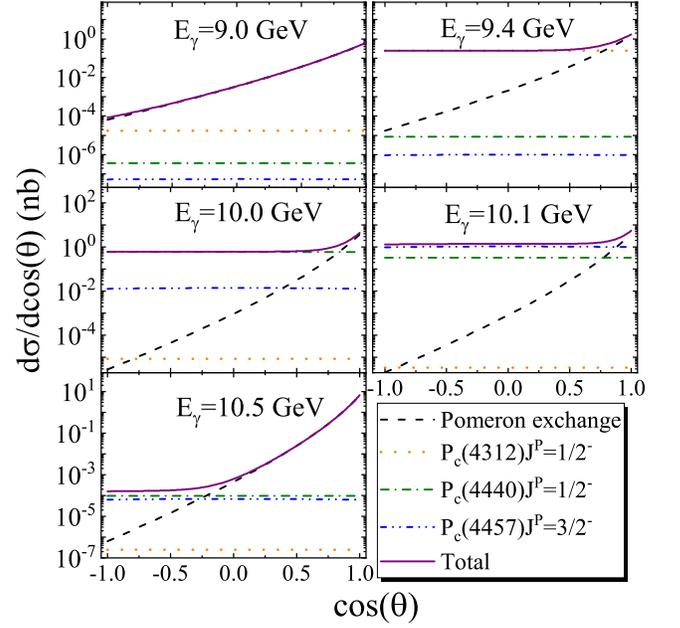}
\caption{The differential cross section $d\protect\sigma %
/d\cos \protect\theta $ of the $\protect\gamma p\rightarrow J/\protect\psi p$
process as a function of $\cos \protect\theta $ at beam energies $E_{\protect%
\gamma }=9.0$, 9.4, 10.0, 10.1, and 10.5 GeV. The coupling constants are
extracted by assuming branching ratio $Br[P_{c}\rightarrow J/\protect\psi %
p]\simeq 3\%$. The black dashed, orange dotted, green dot-dashed, blue
dash-double dotted, and violet solid lines are for the Pomeron exchange, $%
P_{c}(4312),P_{c}(4440),P_{c}(4457)$, and total contributions, respectively.}
\label{dcs01}
\end{figure}

\begin{figure}[h!]
\centering
\includegraphics[scale=0.41]{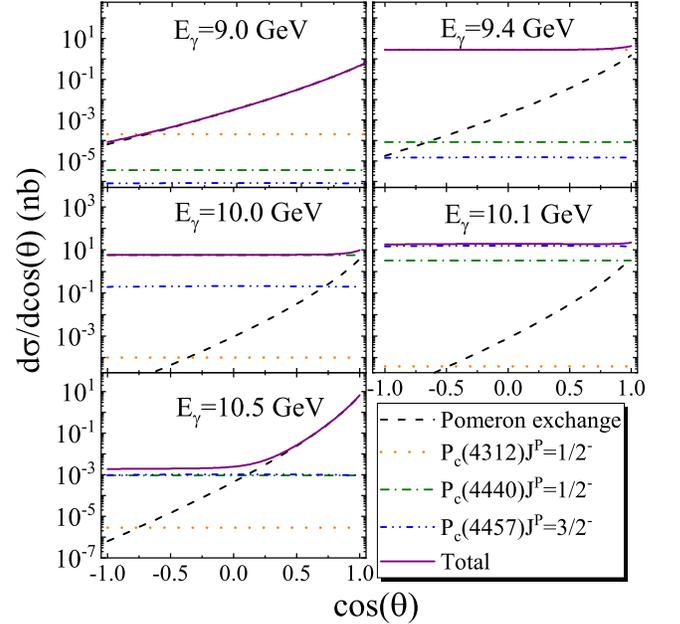}
\caption{Same as Fig. 4 except by assuming branching ratio $%
Br[P_{c}\rightarrow J/\protect\psi p]\simeq 10\%.$}
\label{dcs02}
\end{figure}

The differential cross sections corresponding to branching ratios $%
Br[P_{c}\rightarrow J/\psi p]\simeq 3\%$ and $10\%$ are illustrated in Figs. %
\ref{dcs01} and \ref{dcs02}, respectively.
 One notices that the differential
cross sections are enhanced in the forward direction because of the strong
Pomeron diffractive contribution in the $t$ channel. In the regions of the $%
P_{c}$ states, i.e., $E_{\gamma }=$ 9.4, 10.0, and 10.1 GeV, the effects of
the $P_{c}$ states are obvious, which makes the curve of differential cross
sections tend to be flat.  The  shape  of the differential cross section from the pentaquark  is relevant to the  orbital angular momentum  between the final  and initial particles. Under the assignment of the spin parties of the pentaquarks in the current work,  the pentaquarks can couple to both the final $J/\psi p$ and initial $\gamma p$  in  the S wave and the higher partial waves are neglected because the momentum between the final $J/\psi p$ is fairly small.  The flatness reflects such assignment  of the spin parities. The nondiffractive effects at off-forward angles
in the range of near-threshold energies can be measured by future
experiments, which will help us to clarify the role of $P_{c}$ states in the
 reaction $\gamma p\rightarrow J/\psi p$.


We will now discuss the polarization observables, which can provide crucial
information on the helicity amplitudes and spin structure of a process \cite%
{Fasano:1992es} . To define the polarization observables, the reaction takes
place in the $x-z$ plane. The photon beam asymmetry $\Sigma _{\gamma }$ is
defined as%
\begin{equation}
\Sigma _{\gamma }=\frac{d\sigma (\epsilon _{\perp })-d\sigma (\epsilon
_{\parallel })}{d\sigma (\epsilon _{\perp })+d\sigma (\epsilon _{\parallel })%
},
\end{equation}%
where $\parallel $ and $\perp $ denote the linear polarizations of the
photon along the direction of the $x$ and $y$ axes, respectively. Figure~\ref%
{asymmetry} depicts the numerical results of the photon beam asymmetries $%
\Sigma _{\gamma }$ for the reaction $\gamma p\rightarrow J/\psi p$ at
different beam energies. 
\begin{figure}[h!]
\centering
\includegraphics[scale=0.41]{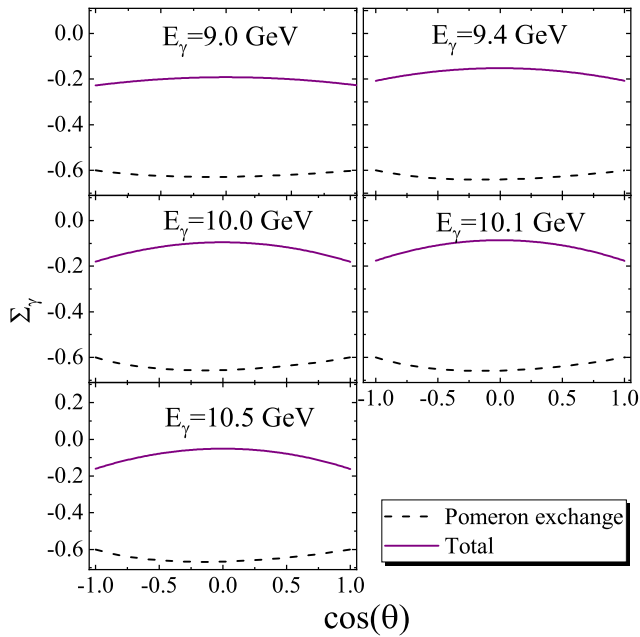}
\caption{The photon beam asymmetries $\Sigma _{\protect\gamma %
}$ for the reaction $\protect\gamma p\rightarrow J/\protect\psi p$ of $E_{%
\protect\gamma }=9.0-10.5$ GeV$.$ The violet solid curves represent the
total results including the $P_{c}$ states, whereas the black dashed lines
only show the result of Pomeron exchange.}
\label{asymmetry}
\end{figure}
One notices that the contributions from the $P_{c}$
states have a greater impact on the polarization observation $\Sigma
_{\gamma }$ near the threshold. Thus, the measurement of the beam asymmetry
will help us clarify the roles of these $P_{c}$ states in the $\gamma
p\rightarrow J/\psi p$ process.

\section{Summary and discussion}

Within the frame of the effective Lagrangian approach and the VMD model, the
production of pentaquark states $P_{c}(4312),P_{c}(4440)$, and $P_{c}(4457)$
via the $s$ channel in the reaction $\gamma p\rightarrow J/\psi p$ is
investigated. Moreover, the $t$-channel Pomeron exchange is also studied,
and is regarded as the background for the photoproduction of the $P_{c}$
states. The numerical results show that existing experimental data for the $%
\gamma p\rightarrow J/\psi p$ process are consistent with the present result
by assuming the branching ratio $Br[P_{c}\rightarrow J/\psi p]\simeq 3\%$.
If the branching ratio of $P_{c}$ decay to $J/\psi p$ is small, these
pentaquark $P_{c}$ states may 
have stronger couplings to other channels, i.e., $\Sigma _{c}\bar{D}$, $%
\Sigma _{c}\bar{D}^{\ast }$ etc. A precision of 0.1 nb/10 MeV is required to
distinguish the $P_{c}(4440)$ and $P_{c}(4457)$. The  total cross
section of the $\gamma p\rightarrow J/\psi p$ process is also calculated by
taking $Br[P_{c}\rightarrow J/\psi p]\simeq 10\%$. The contribution of these
$P_{c}$ states makes several distinct peaks, which are at least 1 order of magnitude
larger than the background cross section, appear in the cross section at
the near-threshold energy region. Hence, if $Br[P_{c}\rightarrow J/\psi
p]\simeq 3\%$ is in line with the actual situation, it is feasible to
search for these pentaquark $P_{c}$ states via the reaction $\gamma p\rightarrow
J/\psi p$ with the precision suggested above, in which the signal can be clearly distinguished from the
background. If the physical  branching ratio is larger, lower  precision will be required in experiment. 

The differential cross sections for the
reaction $\gamma p\rightarrow J/\psi p$ are also calculated. One notices that the cross section of the $t$%
-channel Pomeron exchange is sensitive to the $\theta $ angle and gives a
considerable contribution at forward angles. The contributions of the $P_{c}$
states are mainly concentrated near the threshold energy region and make the
differential cross section relatively flat, which is consistent with our
choice of the spin parities of the pentaquarks. The polarization observable $%
\Sigma _{\gamma }$ is also calculated, and the results suggest that the $P_{c}
$ states have large effects on this observable.

To deepen the understanding of these pentaquark $P_{c}$ states, an
experimental study of the $P_{c}$ states via the photo-induced process is
strongly suggested. The photon beams can be provided at JLab\cite%
{Meziani:2016lhg,Austregesilo:2018mno} and COMPASS \cite{Nerling:2012er}.
The center-of-mass energy 4.5 GeV corresponds to a
laboratory photon energy of 10.5 GeV, which is well within the capabilities
of the GlueX and CLAS12 detectors at JLab \cite%
{Meziani:2016lhg,Austregesilo:2018mno,Kumano:2015gna}. The calculations of
the current work suggest that it is promising to do such experiments at
existing facilities. The expected high-precision data at the threshold energy
region will not only be helpful in clarifying the role of the pentaquark states $%
P_{c}(4312),P_{c}(4440)$, and $P_{c}(4457)$ in
reaction $\gamma p\rightarrow J/\psi p$ but also will help provide important information for better understanding
the nature of these $P_{c}$ states.

\section{Acknowledgments}

This project is supported by the National Natural Science Foundation of
China under Grants No. 11705076 and No. 11675228. We acknowledge the Natural
Science Foundation of Gansu province under Grant No. 17JR5RA113. This work
is partly supported by the HongLiu Support Funds for Excellent Youth Talents
of Lanzhou University of Technology.

\end{document}